\documentclass[aps,twocolumn,superscriptaddress,amsmath,amssymb]{revtex4}

\usepackage{graphicx}

\begin{document}

\title{Power law rheology of generalised Maxwell and Kelvin-Voigt models}

\author{P. Patr\'{\i}cio}
\email{pedro.patricio@adf.isel.pt}
\affiliation{ISEL - Instituto Superior de Engenharia de Lisboa, Instituto Polit\'ecnico de Lisboa, 1959-007 Lisboa, Portugal.}
\affiliation{CFTC - Centro de F\'{i}sica Te\'orica e Computacional, Universidade de Lisboa, Campo Grande, Edif\'{\i}cio C8, P-1749-016 Lisboa}

\date{January 15, 2018}

\begin{abstract}

We analyse the rheological behaviour of the generalised Maxwell and Kelvin-Voigt models with a power law spectrum of relaxation or retardation times, ${\cal H}\sim 1/\tau^{\alpha}$ or ${\cal L}\sim \tau^\alpha$, respectively. When $0<\alpha<1$, both models have the same weak power law behaviour, which occur in a variety of soft materials, including living cells.
Beyond this region, for other values of $\alpha$, 
the two models have different and distinctive rheological behaviours,
converging continuously to the 1-element Maxwell and Kelvin-Voigt models, respectively, 
for $\alpha<-1$ or $\alpha>2$.

\end{abstract}

\maketitle

\section{Introduction}

In a large variety of soft materials, the elastic and viscous moduli 
exhibit approximately the same weak power law behaviour, with the angular frequency of the oscillatory deformations, 
$G'(w)\sim G''(w)\sim w^\alpha$ ($0<\alpha<1$).
In diluted unentangled polymers, the power law exponent is associated with
the relaxation modes of each molecular chain \cite{rouse1953theory,zimm1956dynamics,doi1988theory,rubinstein2003polymers}.
In pre-gel and post-gel polymers, it is associated with percolation and network self-similarity
\cite{de1976relation,cates1985brownian,muthukumar1985dynamics,
winter1986analysis,chambon1987linear,muthukumar1989screening,goldenfeld1992dynamic,tanaka2011polymer} (an extensive review is found in \cite{winter1997rheology}).
The universality of the weak power law behaviour, even for systems devoid of any obvious fractal structure, like ``foams, emulsions, pastes and slurries",
led Sollich and coworkers to propose the phenomenological ``Soft Glassy Materials'' (SGM) model \cite{sollich1997rheology,sollich1998rheological}.
Based on the idea of structural disorder and metastability, this model relates the power law exponent to a mean-field noise temperature, closely above the glass transition.

In the context of cell mechanics, the weak power law behaviour of the cytoskeleton has been recently associated with
the ``Glassy Worm Like Chain" model \cite{kroy2007glassy,wolff2010inelastic}, whose relaxation modes
follow essentially the ``Worm Like Chain" model for semiflexible polymers, of which the cytoskeleton is composed, but retain nevertheless some analogy with the SGM model (see \cite{pritchard2014mechanics} for a review).

\begin{figure}[htp]
\begin{center}
\begin{tabular}{cc}
\includegraphics[scale=0.4]{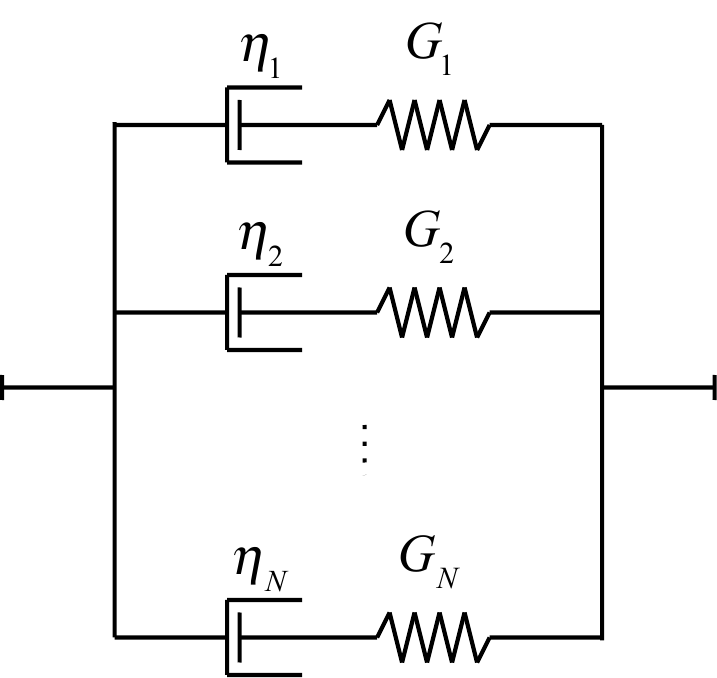}\\
\includegraphics[scale=0.4]{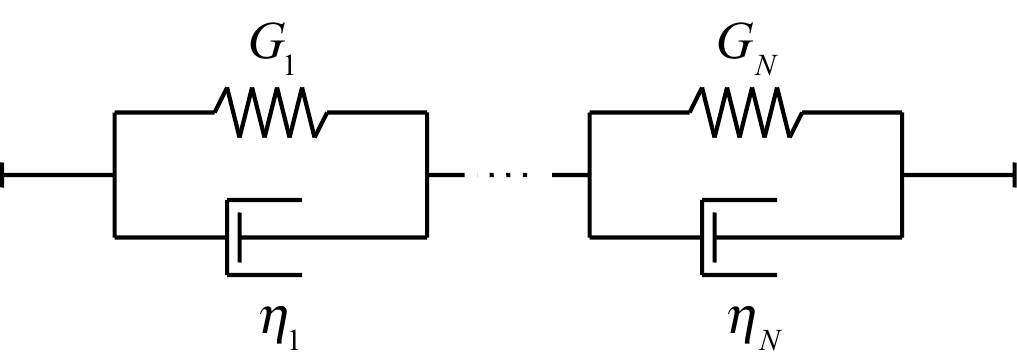}\\
\end{tabular}
\caption{Generalised Maxwell (top) and Kelvin-Voigt (bottom) models.
The generalised Maxwell model is composed of a collection of Maxwell elements in parallel, each made out ($0<\alpha<1$) of a dashpot of drag coefficient $\eta_i$ in series with a spring of stiffness $G_i$. Each Maxwell element corresponds to a different relaxation mode, with relaxation time $\tau_i=\eta_i/G_i$.
The generalised Kelvin-Voigt model is composed of a collection of Kelvin-Voigt elements in series, 
each made out of a dashpot in parallel with a spring. Each Kelvin-Voigt element corresponds to a different retardation mode, 
with retardation time $\tau_i=\eta_i/G_i$.}
\label{fig1}
\end{center}
\end{figure}

To understand this behaviour, 
the different models use a microscopic description of the materials 
to propose a specific spectrum of relaxation or retardation modes. 
Thus, they may be directly associated with the generalised Maxwell or Kelvin-Voigt models (see Fig. \ref{fig1})
from the linear theory of viscoelasticity \cite{christensen2012theory,phan2012understanding}.
In fact, it is known that we may mathematically translate the relaxation spectrum of the generalised Maxwell model to the retardation spectrum of the generalised Kelvin-Voigt model.
The general relation is however non trivial and, in some situations, a spring and/or a dashpot should be added, in parallel for the generalised Maxwell model 
and in series for the generalised Kelvin-Voigt model \cite{lakes1998viscoelastic}.

In this article, we will firstly highlight the fundamental aspects of these interesting and universal weak power law behaviours, in the framework of the linear theory of viscoelasticity.
In the following sections, we will make a parallel excursion to the rheological behaviour of the generalised Maxwell and Kelvin-Voigt models, with a power law spectrum of relaxation or retardation times, ${\cal H}\sim 1/\tau^{\alpha}$ or ${\cal L}\sim \tau^\alpha$, respectively. When $0<\alpha<1$, we will see that these models lead to approximately the same weak power law behaviour, so often encountered in nature. For other values of $\alpha$, 
the two models have different and distinctive rheological behaviours,
converging continuously to the 1-element Maxwell and Kelvin-Voigt models, respectively, 
for $\alpha<-1$ or $\alpha>2$.

\section{Weak power law rheology}
\label{sec: Power law rheology}

In the linear theory of viscoelasticity \cite{phan2012understanding}, the relaxation function $G(t)$ is defined through the force response
\begin{equation}
\sigma(t)=\varepsilon_0 G(t)
\label{Eq: Relaxation function}
\end{equation}
to an applied step deformation $\varepsilon=\varepsilon_0 \Theta(t)$ ($\Theta(t)$ is the step function).
Analogously, the creep compliance $J(t)$ is defined through the deformation response 
\begin{equation}
\varepsilon(t)=\sigma_0J(t)
\label{Eq: Creep compliance}
\end{equation}
to an applied step force $\sigma=\sigma_0 \Theta(t)$.

More generally, if we impose an arbitrary deformation $\varepsilon(t)$, which may be viewed as an infinite sum of small successive step deformations $d\varepsilon(\tau)$, with $0<\tau<t$, then Boltzmann superposition principle states that the total force exerted on the system may be determined by:
\begin{equation}
\sigma(t)=\int_0^t G(t-\tau)d\varepsilon(\tau)=\int_0^t G(t-\tau) \frac{d\varepsilon(\tau)}{d\tau}d\tau.
\label{Eq: Boltzmann superposition for G}
\end{equation}
Similarly, for an imposed general force $\sigma(t)$, the total deformation of the system is given by:
\begin{equation}
\varepsilon(t)=\int_0^t J(t-\tau) \frac{d\sigma(\tau)}{d\tau}d\tau.
\label{Eq: Boltzmann superposition for J}
\end{equation}

If the system is subjected to an oscillatory deformation 
$\varepsilon(t)=\varepsilon_0e^{iwt}$ or force $\sigma(t)=\sigma_0e^{iwt}$, two important rheological assays, its response is given, respectively, by
\begin{equation}
\sigma(t)=\varepsilon_0G^*e^{iwt},
\label{Eq: Complex modulus}
\end{equation}
where  $G^*=G'+iG''$ is the complex modulus, or, 
\begin{equation}
\varepsilon(t)=\sigma_0J^*e^{iwt},
\label{Eq: Complex compliance}
\end{equation}
where $J^*=J'-iJ''$ is the complex compliance.
$G'$ ($J'$) and $G''$ ($J''$) are defined as the elastic and viscous moduli (compliances).

The Boltzmann superposition principle, expressed by Eq. (\ref{Eq: Boltzmann superposition for G}) 
and (\ref{Eq: Boltzmann superposition for J}), allows us to relate the different rheological functions through:
\begin{equation}
G^*(w)=(iw)\tilde G(iw)=\frac{1}{(iw)\tilde J(iw)}=\frac{1}{J^*(w)}
\label{Eq: Viscoelastic relations}
\end{equation}
where the Laplace transform is given by $\tilde f(s)=\int_0^\infty f(t) e^{-st} dt$.

From these relations it may be made clear that any relaxation function with a weak power law behaviour:
\begin{equation}
G(t)\sim\frac{1}{t^\alpha}\;\;\;\;(0<\alpha<1)
\label{Eq: Weak power law for relaxation}
\end{equation}
will necessarily lead to the weak power law behaviours of the elastic and viscous moduli.
Indeed, the relaxation function is necessarily a decreasing function of time, implying $\alpha>0$.
For $\alpha<1$, its Laplace transform is given by $\tilde G(s)\sim s^{\alpha-1}$, the complex modulus becomes $G^*(w)\sim(iw)^\alpha=w^\alpha e^{i\pi\alpha/2}$, yielding
\begin{equation}
G'(w)\sim G''(w)\sim w^\alpha,\;\;\;\;\frac{G''}{G'}=\tan\frac{\pi \alpha}{2}
\label{Eq: Weak power law for complex modulus}
\end{equation}
It should be noted that, in these conditions ($0<\alpha<1$), 
$\tilde J(s)\sim 1/s^{\alpha+1}$ and the creep compliance, which is necessarily an increasing function of time, is given by
\begin{equation}
J(t)\sim t^\alpha.
\label{Eq: Weak power law for creep compliance}
\end{equation}

These are the fundamental features of the weak power law rheology. A system with this type of behaviour is considered to be solid like whenever $0<\alpha<0.5$ (because $G''/G'<1$) and fluid like whenever $0.5<\alpha<1$ (as $G''/G'>1$).

In the following sections, we will see that these weak power law behaviours, with ($0<\alpha<1$), are found in systems described by the generalised Maxwell or Kelvin-Voigt models with a power law spectrum of relaxation or retardation times, ${\cal H}\sim 1/\tau^{\alpha}$ or ${\cal L}\sim \tau^\alpha$ , respectively.
These spectra are however well defined for all values of $\alpha$. For $\alpha<0$, or $\alpha>1$, we will see 
that these two two models have different and distinctive rheological behaviours, 
converging continuously to the 1-element Maxwell and Kelvin-Voigt models, respectively, 
for $\alpha<-1$ or $\alpha>2$.

\section{Generalised Maxwell model}

We consider first a generalised Maxwell model (see Fig. \ref{fig1}, top) composed of a parallel array of $N$ Maxwell viscoelastic elements.
Each Maxwell element is composed of a dashpot and a spring in series.
These are the basic elements of the linear theory of viscoelasticity. A spring of stiffness $G$ follows Hooke's law, $\sigma=G\varepsilon$, and a 
dashpot of drag coefficient $\eta$ follows Newton's law, $\sigma=\eta \dot \varepsilon=\eta d\varepsilon/dt$.
As a result, the Maxwell element behaves elastically on small times scales and as a fluid on long ones.
Its relaxation function is $G(t)=Ge^{-t/\tau}$, where $\tau=\eta/G$ is the element relaxation time.

Because the Maxwell elements are in parallel, the total relaxation function is just the sum:
\begin{equation}
G(t)=\sum_{i=1}^{N}G_ie^{-t/\tau_i}
\label{Eq: GM relaxation function}
\end{equation}
This sum may be approximated by an integral by multiplying it by $di=1$. If $di$ is considered to be small, we have:
\begin{equation}
G(t)\approx \int_{\tau_1}^{\tau_N}G_\tau e^{-t/\tau}\frac{di}{d\tau}d\tau
\end{equation}
The derivative $di/d\tau$ corresponds to the density of modes (per unit relaxation time). Choosing the change of variables $\zeta=1/\tau$,
and for $t$ between the minimum and maximum relaxation times, $\tau_{\min}$ and $\tau_{\max}$,
this integral approaches a Laplace transform. To establish the weak power law behaviour
$G(t)\sim1/t^\alpha$ (with $0<\alpha<1$), the relaxation spectrum must scale as
\begin{equation}
{\cal H}(\tau)\equiv G_\tau\frac{di}{d\tau}\tau\sim \frac{1}{\tau^{\alpha}}
\label{Eq: Weak power law relaxation spectrum}
\end{equation}

As with other rheological functions, 
the relaxation spectrum of a viscoelastic fluid characterises its rheological behaviour. 
Because ${\cal H}(\tau)$ is defined in terms of an arbitrary choice of stiffnesses, $G_\tau$, 
and density of modes, $di/d\tau$, the relaxation spectrum is not subjected to any constraint, and
$\alpha$ may assume any value.

To understand the rheological behaviour of a 
material with a general power law spectrum of relaxation times,
we will determine the corresponding elastic and viscous moduli.
After performing a Laplace transform and using relations (\ref{Eq: Viscoelastic relations}), 
we have
\begin{align}
\label{Eq: GM elastic modulus}
G'(w)&\sim \int_{\tau_{\min}}^{\tau_{\max}}\frac{1}{\tau^\alpha}\frac{(w\tau)^2}{1+(w\tau)^2}\frac{d\tau}{\tau}\\
G''(w)&\sim \int_{\tau_{\min}}^{\tau_{\max}}\frac{1}{\tau^\alpha}\frac{w\tau}{1+(w\tau)^2}\frac{d\tau}{\tau}.
\label{Eq: GM viscous modulus}
\end{align}

For $w\ll 1/\tau_{\max}$, the integrals (\ref{Eq: GM elastic modulus}) and (\ref{Eq: GM viscous modulus})
are much simplified, and we obtain $G'\sim w^2$ and $G''\sim w^1$. For $w\gg 1/\tau_{\min}$, 
we have simply $G'\sim w^0$ and $G''\sim w^{-1}$. These scalings correspond to the scalings of the 1-element Maxwell model, both for small and large $w$.

For $1/\tau_{\max}\ll w \ll 1/\tau_{\min}$, 
we may extend the limits of the integrals (\ref{Eq: GM elastic modulus}) and (\ref{Eq: GM viscous modulus}) to 
$\tau_{\min}\to 0$ and $\tau_{\max}\to \infty$, which allow us to obtain the results:
\begin{align}
G'(w)&\sim \frac{\pi}{2}w^\alpha\csc\frac{\pi\alpha}{2}\;\;\;(\text{if}\;\;0<\alpha<2)\\
G''(w)&\sim \frac{\pi}{2}w^\alpha\sec\frac{\pi\alpha}{2}\;\;\;(\text{if}\;\;-1<\alpha<1).
\end{align}

When $0<\alpha<1$, both integrals are well defined and we recover the weak power law behaviours
of Eq. (\ref{Eq: Weak power law for complex modulus}), as expected.
If $\alpha<0$. the integral (\ref{Eq: GM elastic modulus}) for the elastic modulus diverges as $\tau_{\min}\to 0$.
The elastic modulus is then dominated by the smallest relaxation time, and we have $G'\sim w^0$.
On the contrary, if $\alpha>2$, then this integral is dominated by the largest relaxation time, and we have $G'\sim w^2$. By the same line of reasoning, we may determine from Eq. (\ref{Eq: GM viscous modulus}) the power 
law behaviours $G''\sim w^{-1}$ for $\alpha<-1$ and $G''\sim w^1$ for $\alpha>1$.

Applying these results to each interval of $\alpha$, we obtain the interesting power law behaviours:
\begin{align}
&G'\sim w^0 & &G''\sim w^{-1} & &(\alpha<-1) \\
&G'\sim w^0 & &G''\sim w^{\alpha} & &(-1<\alpha<0) \\
&G'\sim w^\alpha & &G''\sim w^{\alpha} & &(0<\alpha<1) \\
&G'\sim w^{\alpha} & &G''\sim w^1 & &(1<\alpha<2) \\
&G'\sim w^2 & &G''\sim w^1 & &(2<\alpha)
\end{align}

We note that for $\alpha<-1$ or $\alpha>2$, the rheological response of the material is entirely
dominated by only one Maxwell element, corresponding respectively to the maximum or minimum relaxation times.

To plot representative elastic and viscous moduli vs $w$, for different values of $\alpha$, 
we used the discrete generalised Maxwell model, with well defined $N$ relaxation modes.
The power law relaxation spectrum of Eq. (\ref{Eq: GM relaxation function}) 
may be implemented from two independent functions, $G_\tau$ and $di/d\tau$. 
So, one of them may be chosen arbitrarily.

\begin{figure}[htp]
\begin{center}
\begin{tabular}{cc}
\includegraphics[scale=0.17]{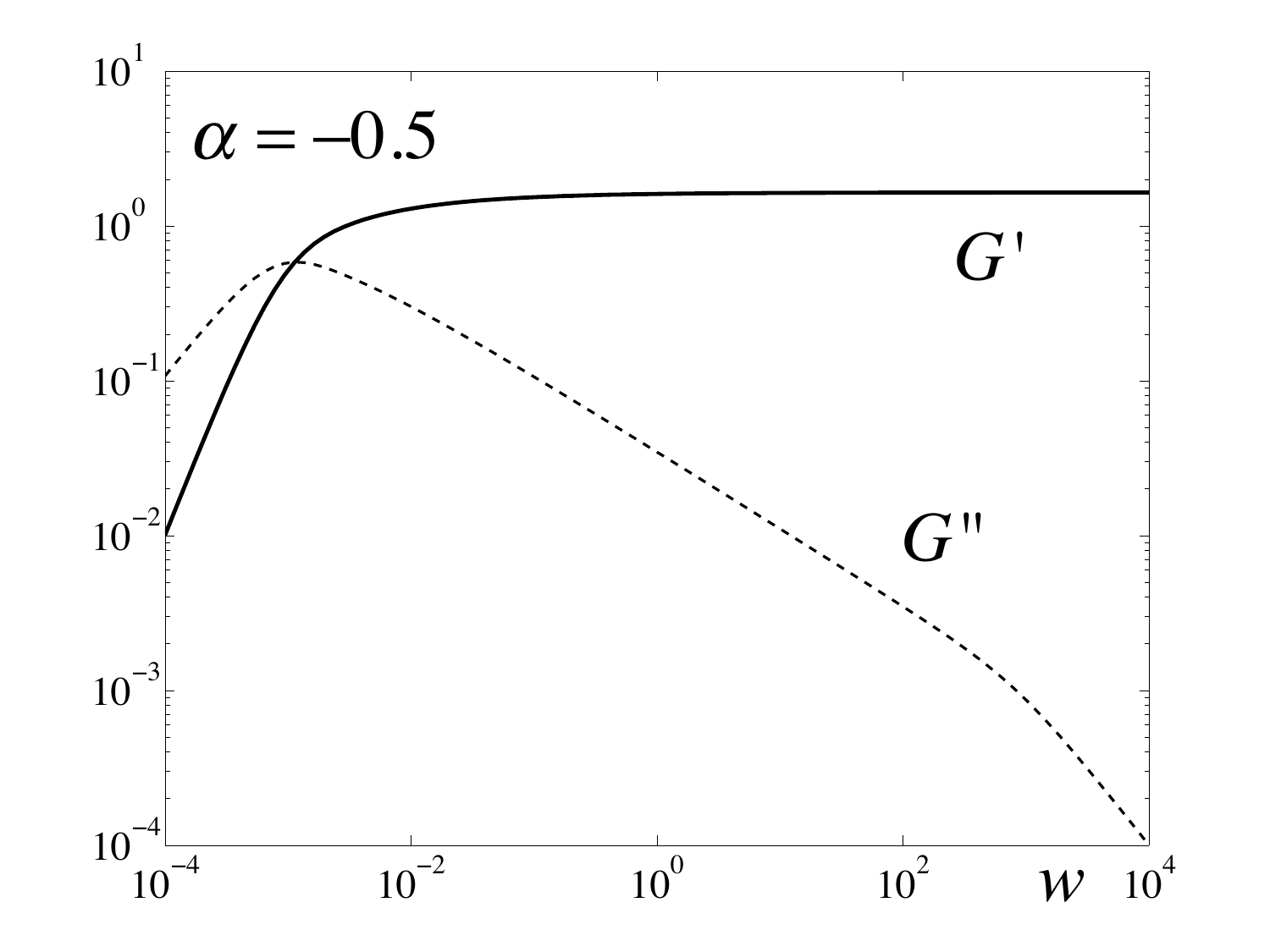}\includegraphics[scale=0.17]{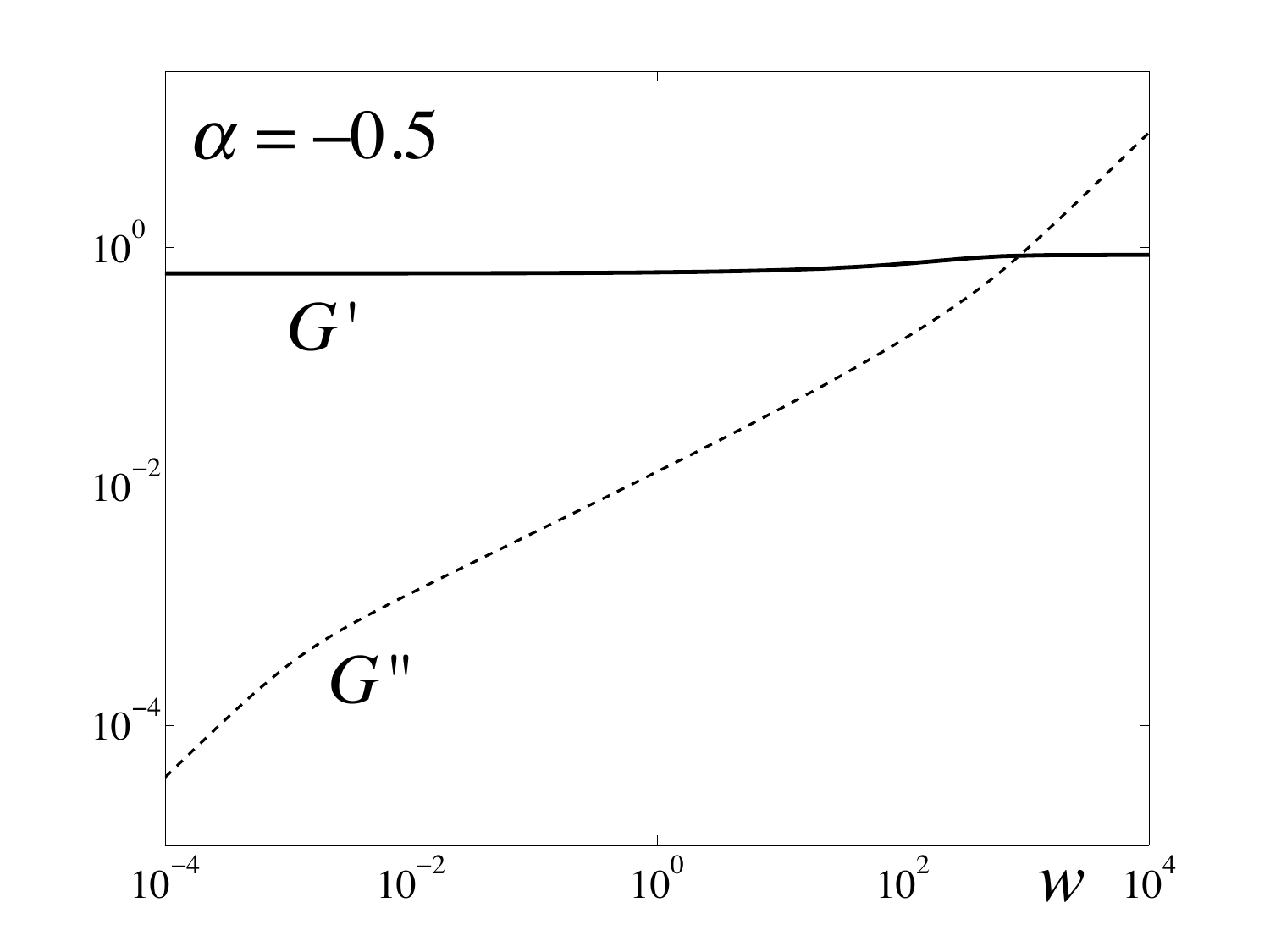}\\
\includegraphics[scale=0.17]{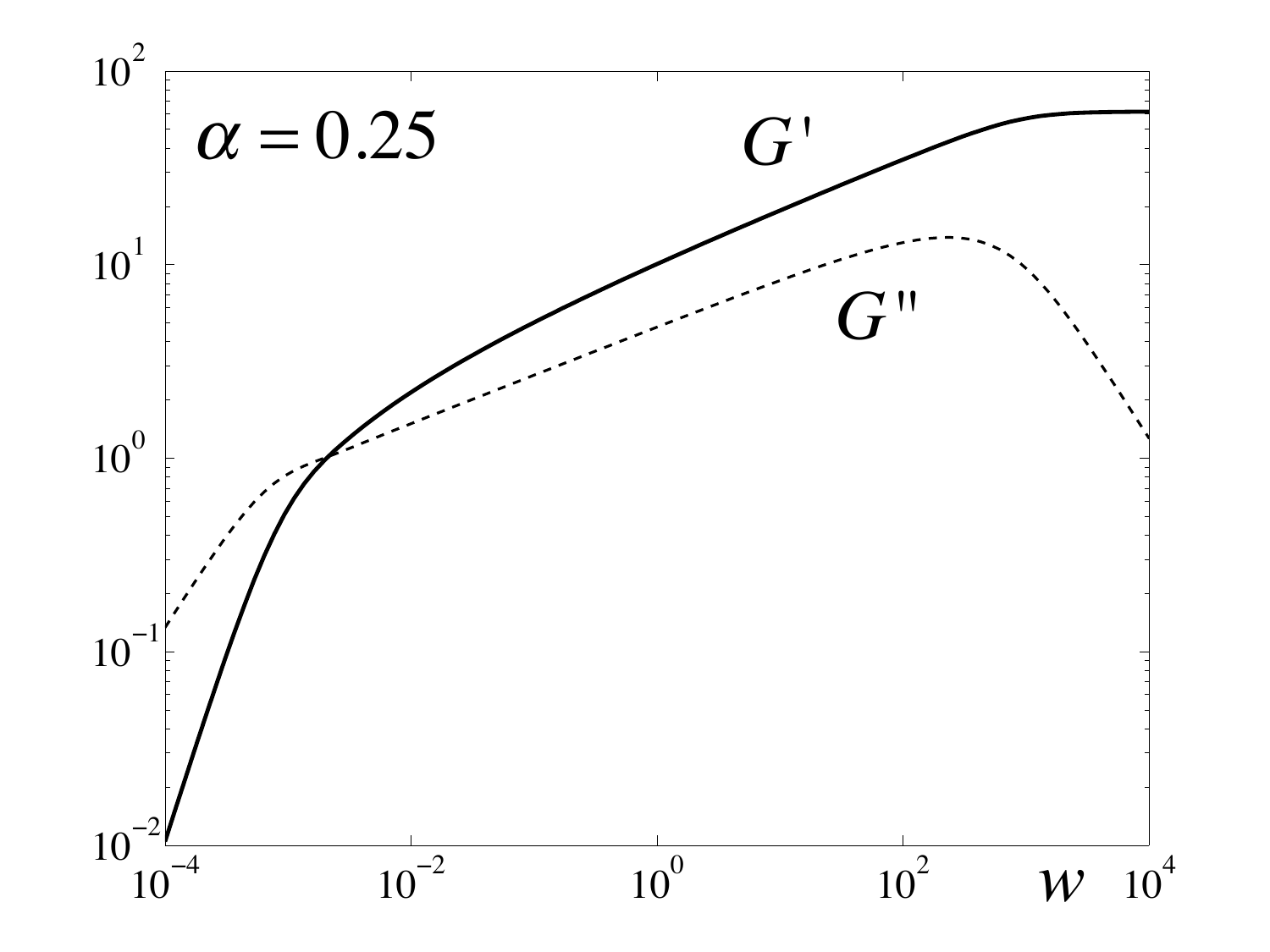}\includegraphics[scale=0.17]{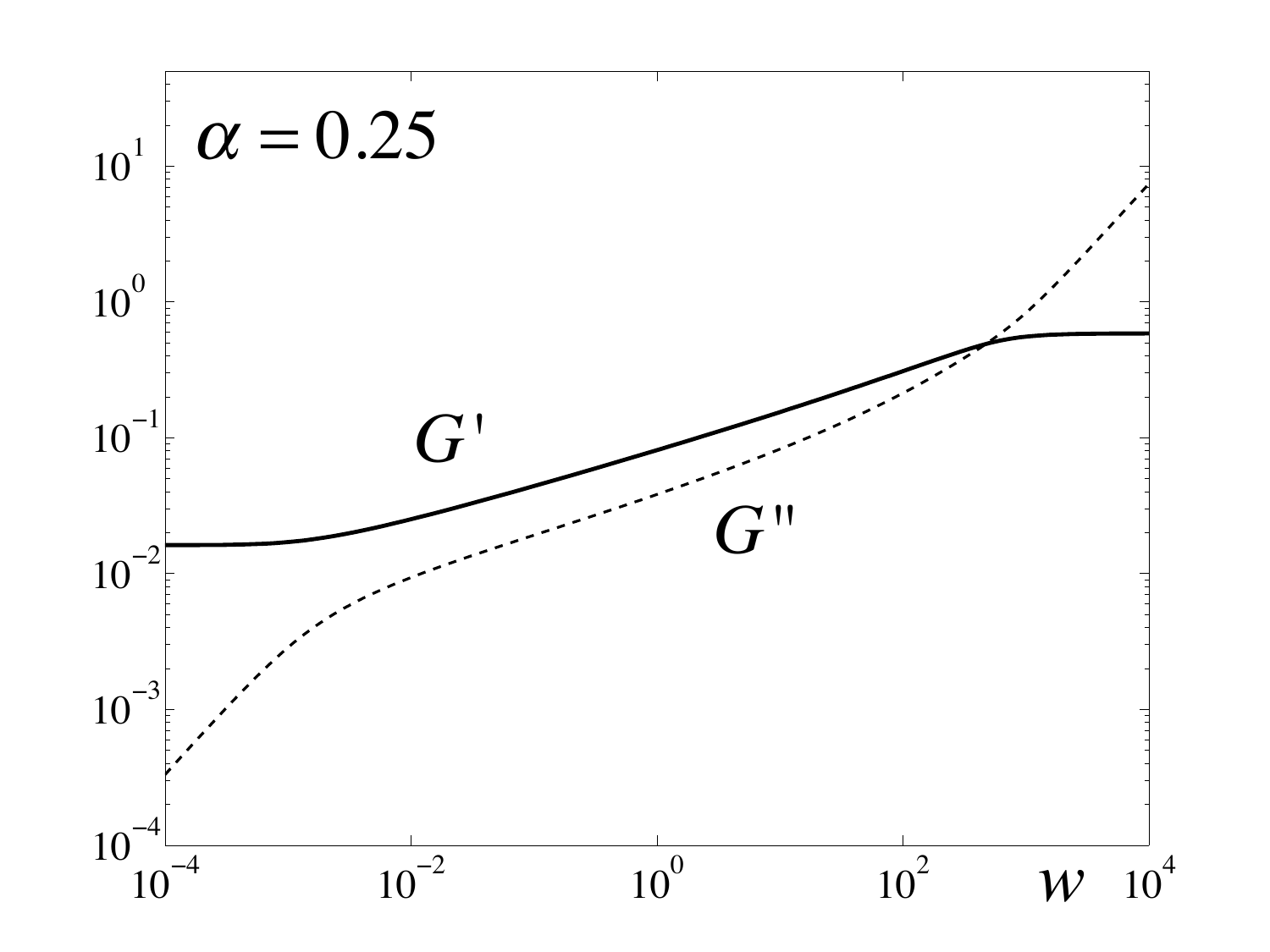}\\
\includegraphics[scale=0.17]{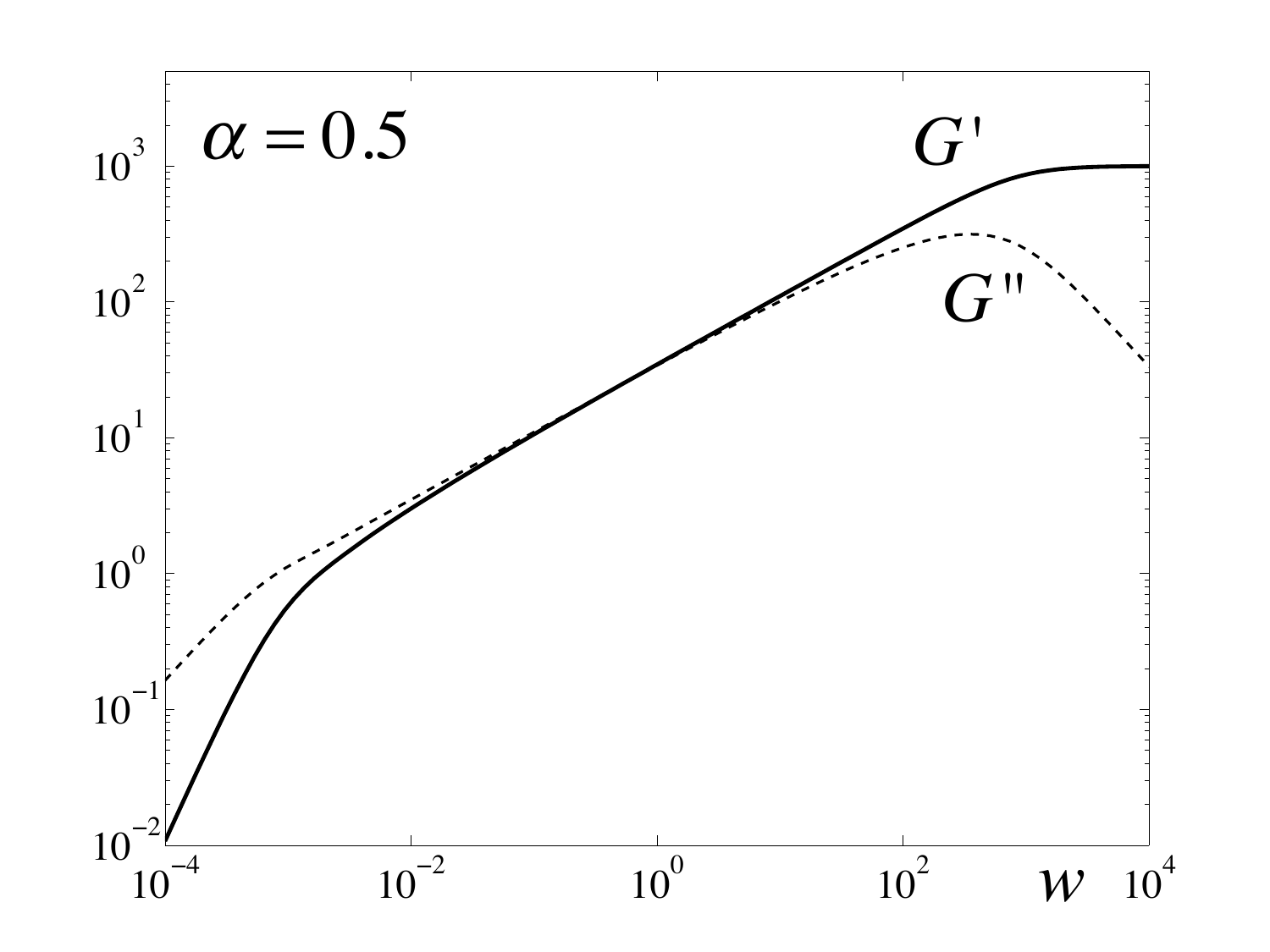}\includegraphics[scale=0.17]{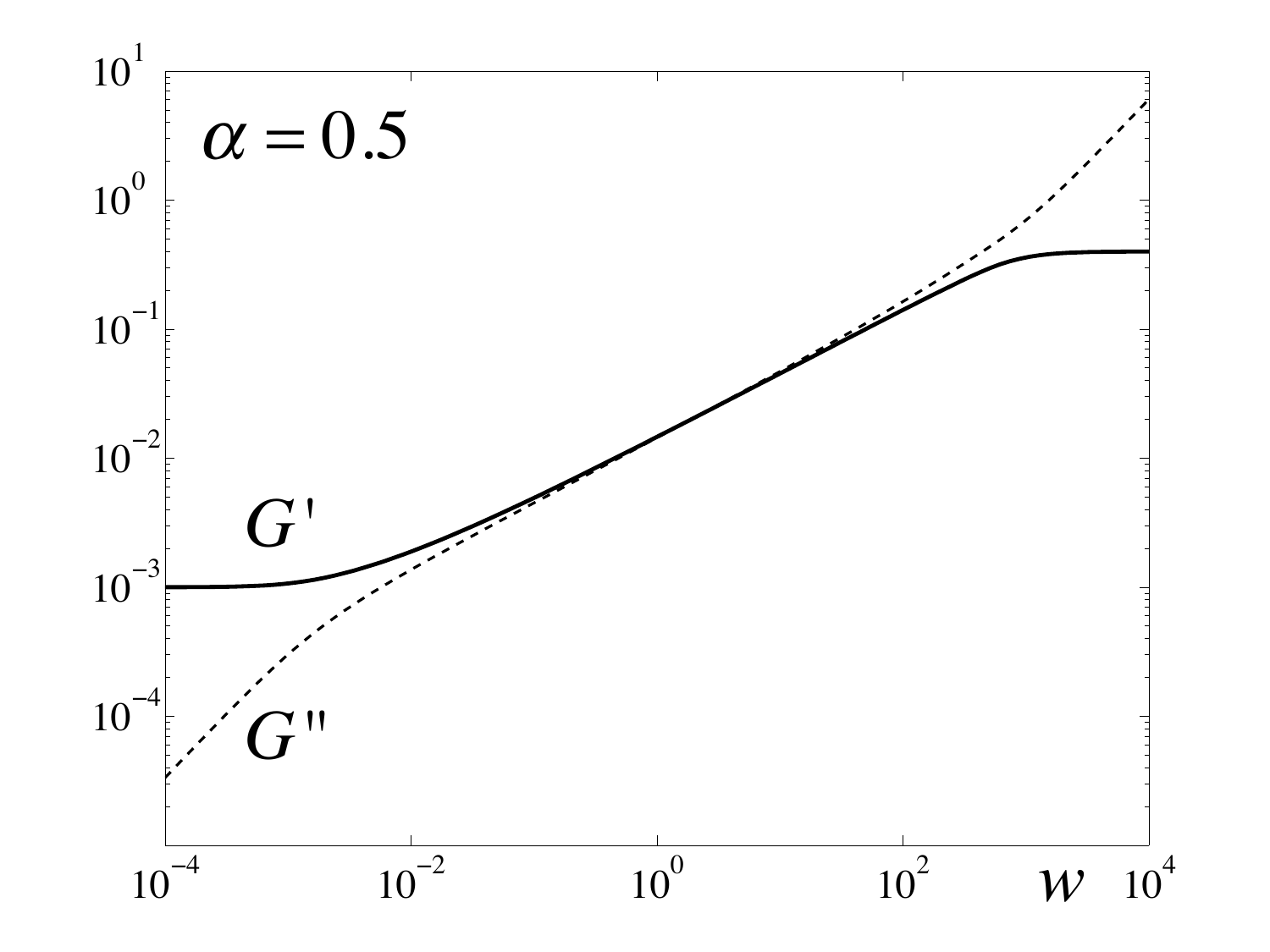}\\
\includegraphics[scale=0.17]{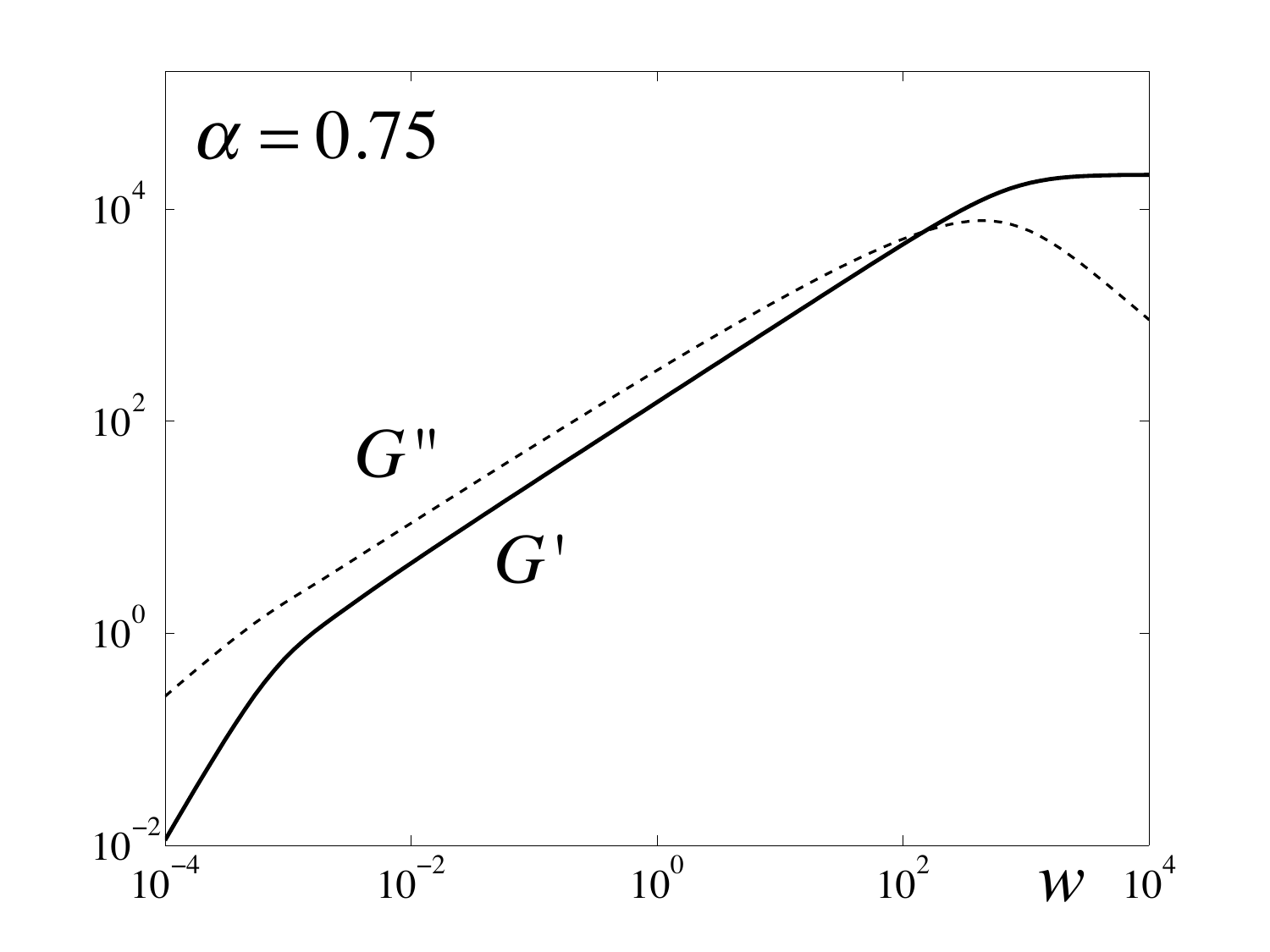}\includegraphics[scale=0.17]{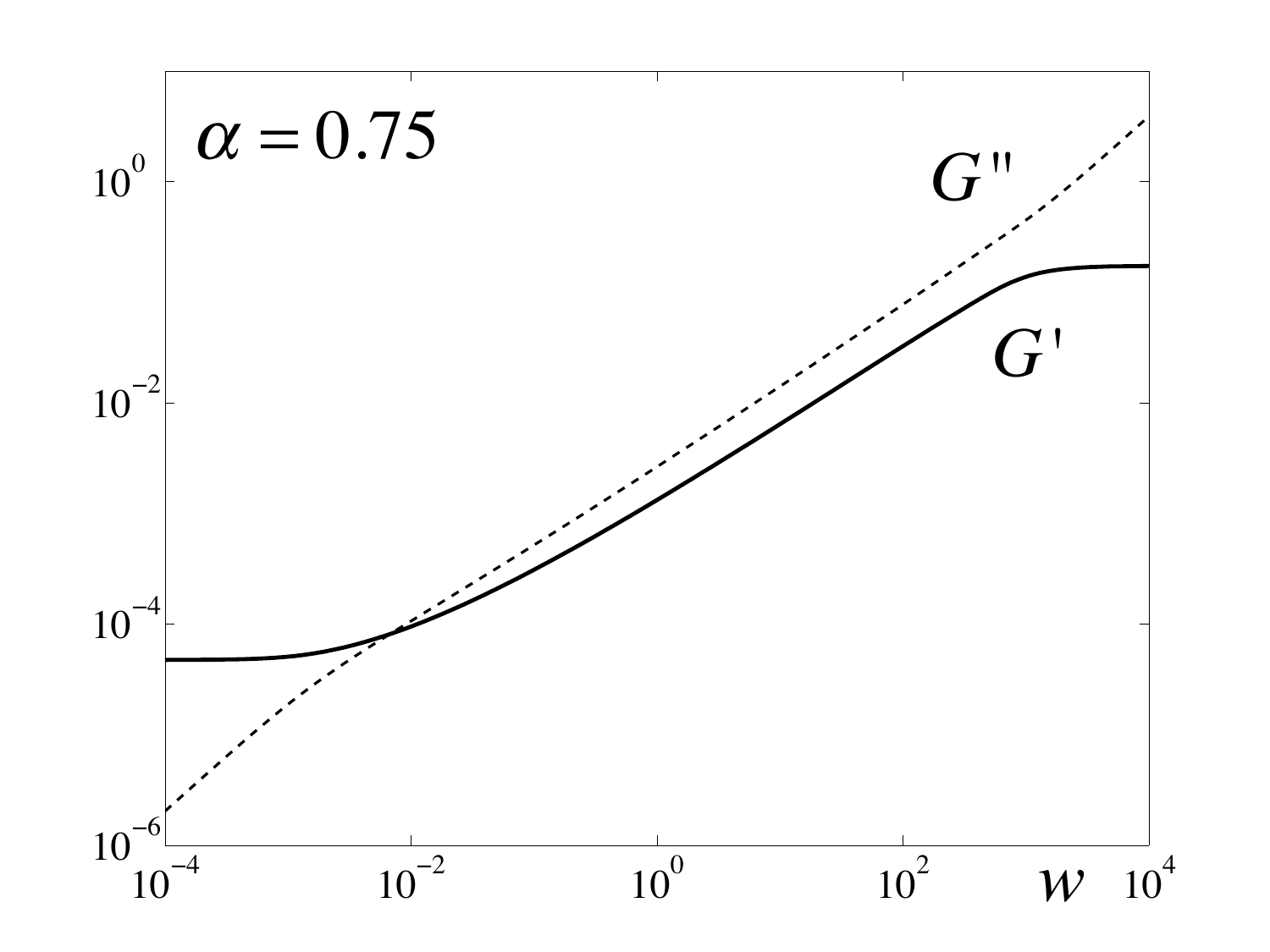}\\
\includegraphics[scale=0.17]{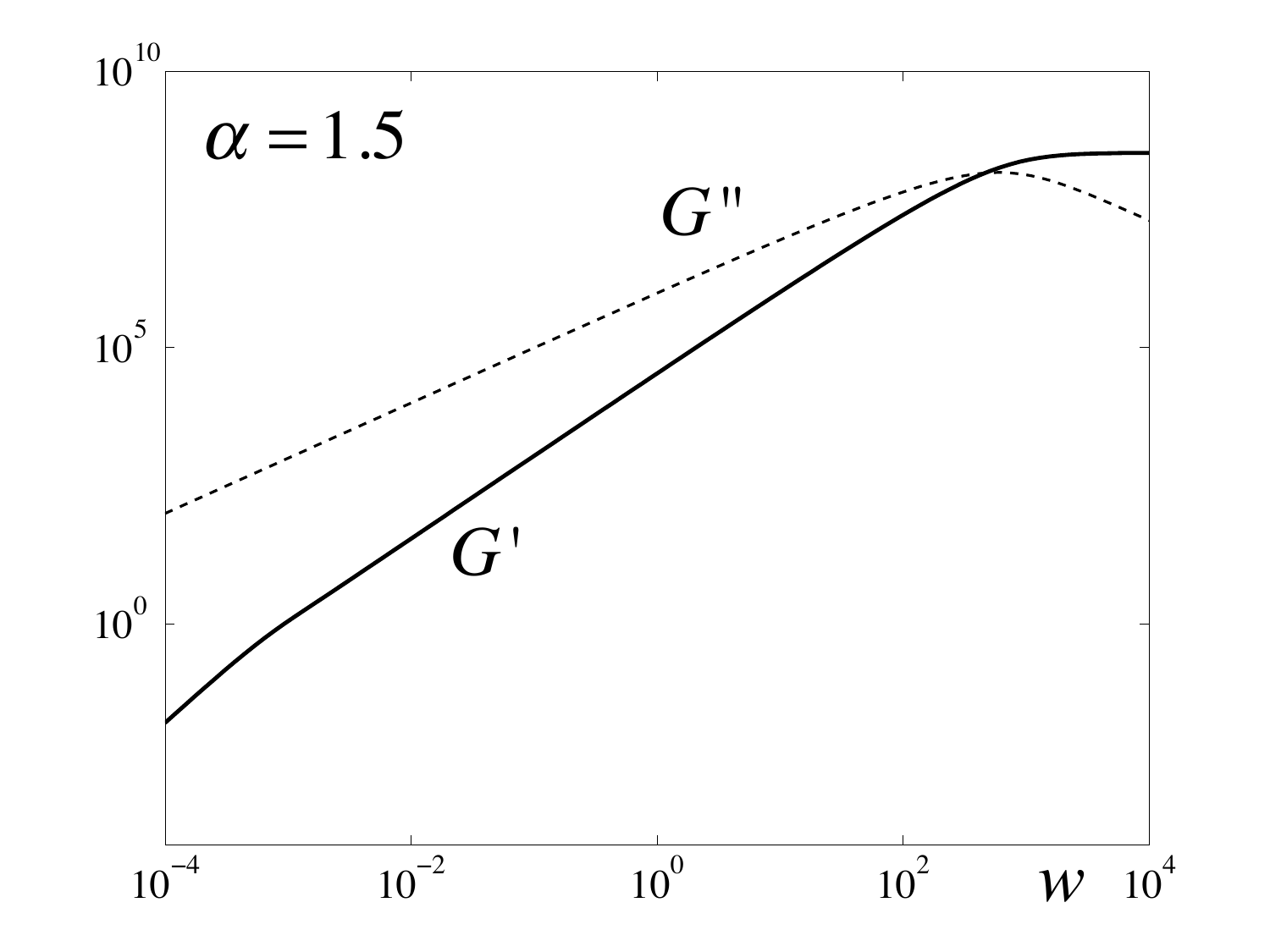}\includegraphics[scale=0.17]{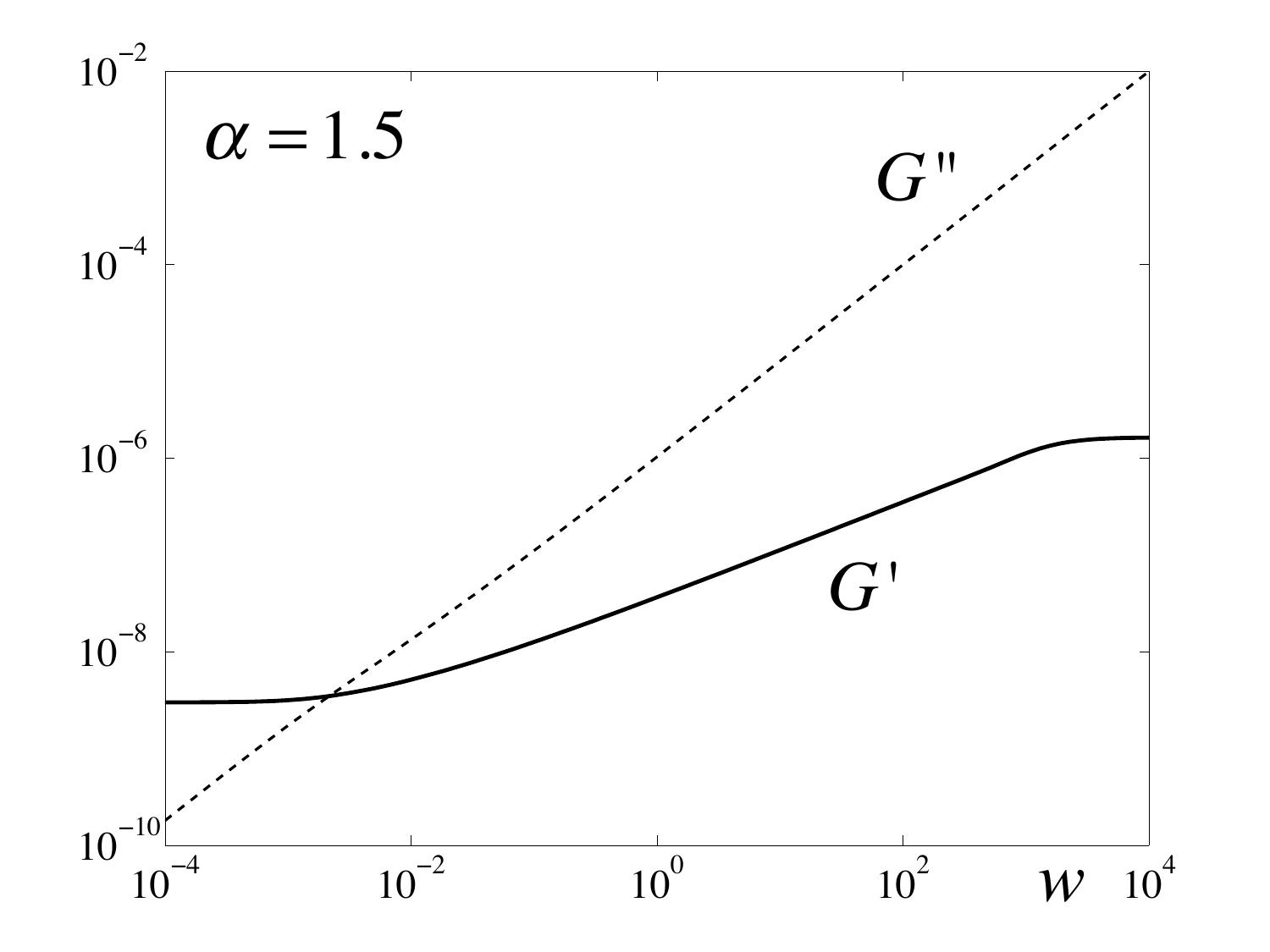}
\end{tabular}
\caption{Elastic and viscous moduli, $G'(w)$ (solid line) and $G''(w)$ (dashed line) (in units of $G$)
obtained from the generalised Maxwell relaxation spectrum of Eq. (\ref{Eq: General GM relaxation spectrum}) (left) and
the generalised Kelvin-Voigt retardation spectrum of Eq. (\ref{Eq: General GK retardation spectrum}) (right),
for $\alpha=-0.5,0.25,0.5,0.75,1.5$ (top to bottom), $\tau_{\min}=10^{-3}$, $\tau_{\max}=10^3$ (in arbitrary units), and $N=10^3$.}
\label{fig2}
\end{center}
\end{figure}

An interesting choice consists of assuming the power law behaviour $G_i=Gi^\beta$ for the stiffnesses.
In this case, the power law spectrum of relaxation times is recovered through:
\begin{equation}
\tau_i=\frac{\tau_{1}}{i^{(1+\beta)/\alpha}}\;\;\;\;(G_i=Gi^\beta),
\label{Eq: General GM relaxation spectrum}
\end{equation}
independently of our choice of $\beta$ (the density of modes $di/d\tau$ is easily obtained from Eq. (\ref{Eq: General GM relaxation spectrum})). 

Mathematically, this choice allows us to define not only the time relaxation limits, 
$\tau_{\max}$, $\tau_{\min}$, but also the exact number of elements $N$ we want to use, for any $\alpha$.
We just have to take $\tau_1=\tau_{\max}$ and $\beta=( \alpha \ln (\tau_{\max}/\tau_{\min})/\ln N)-1$.

If we use the Laplace transform of Eq. (\ref{Eq: GM relaxation function}), we obtain,  without approximation,
the well-known elastic and viscous moduli of the generalised Maxwell model, for any choice of discrete relaxation spectra:
\begin{align}
\label{Eq: G' sums}
G'(w)&=\sum_{i=1}^{N}G_i\frac{(w\tau_i)^2}{1+(w\tau_i)^2}\\
G''(w)&=\sum_{i=1}^{N}G_i\frac{w\tau_i}{1+(w\tau_i)^2}.
\label{Eq: G'' sums}
\end{align}

Figure \ref{fig2} (left) shows the elastic and viscous moduli, $G'(w)$ and $G'(w)$ for 
$\alpha=-0.5,0.25,0.5,0.75,1.5$, $\tau_{\min}=10^{-3}$, $\tau_{\max}=10^3$ (in arbitrary units), and $N=10^3$.
To obtain these plots, we have used the sums of Eq. (\ref{Eq: G' sums}) and (\ref{Eq: G'' sums}), together with
the discrete relaxation spectrum defined by Eq. (\ref{Eq: General GM relaxation spectrum}).

\begin{figure}[htp]
\begin{center}
\begin{tabular}{cc}
\includegraphics[scale=0.32]{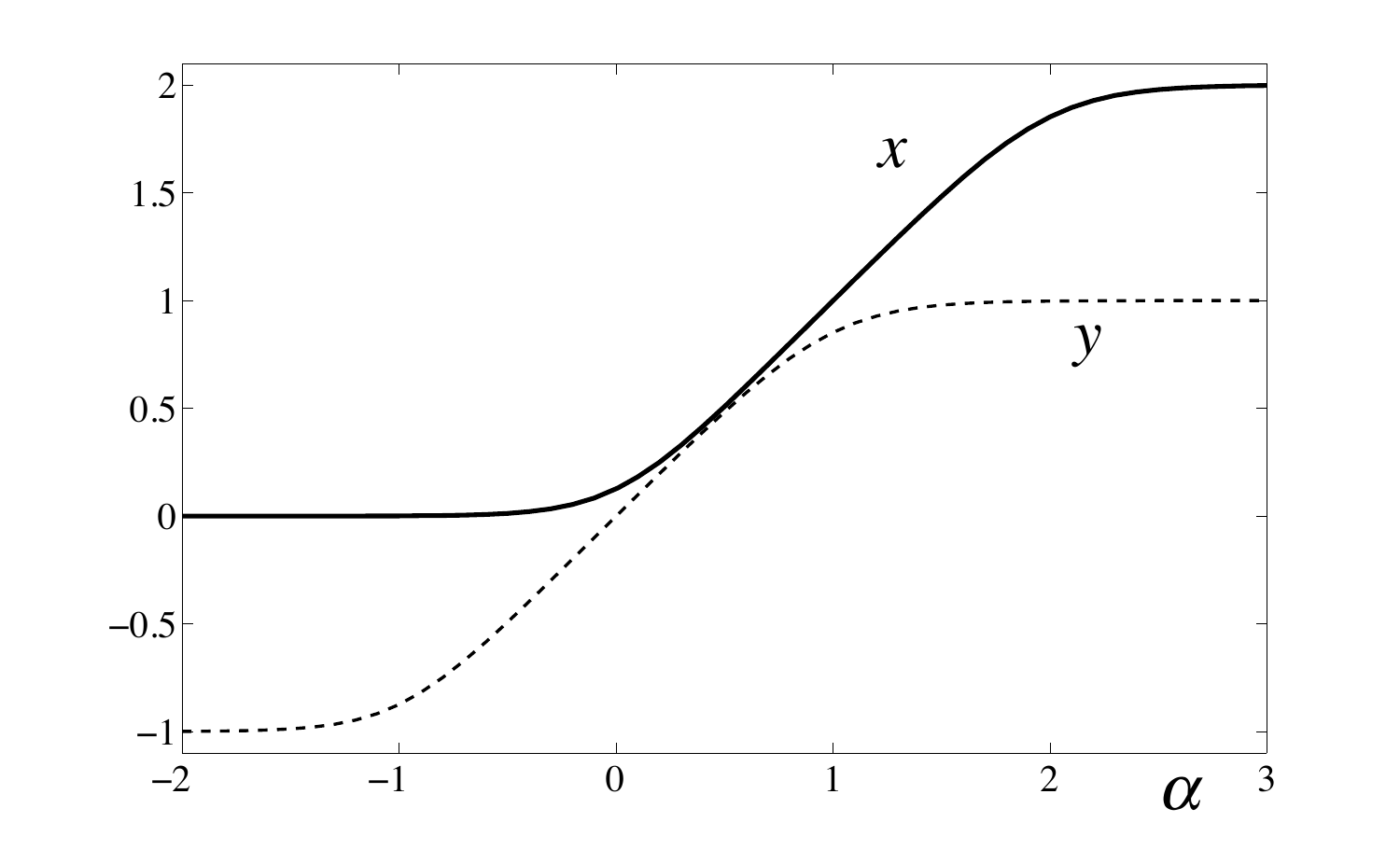}\\
\includegraphics[scale=0.32]{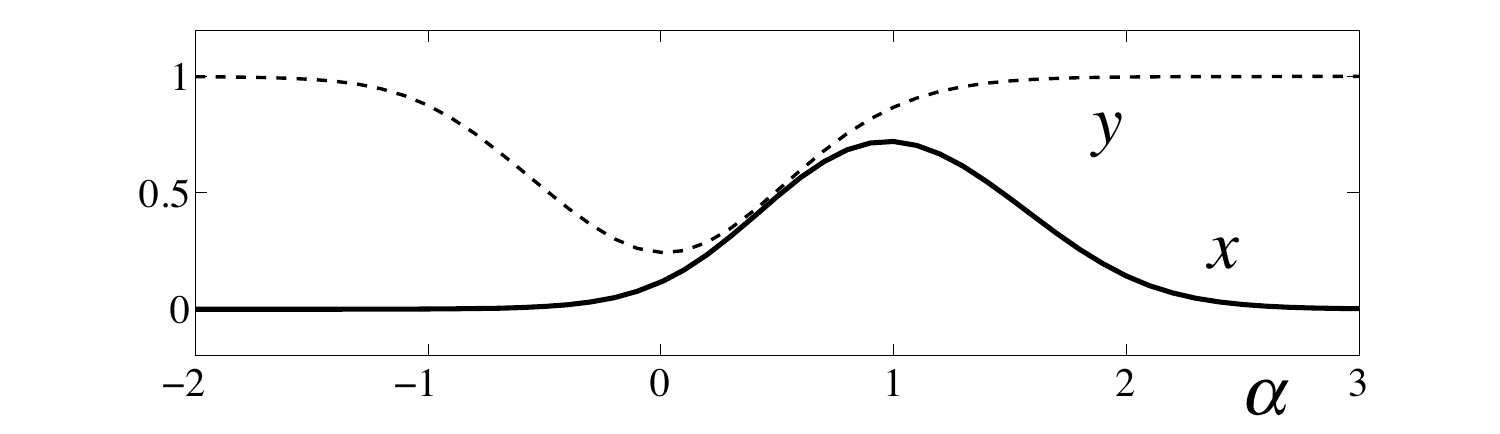}\\
\end{tabular}
\caption{Exponents $x$ and $y$ vs $\alpha$ of the power laws $G'\sim w^x$ and $G''\sim w^y$, obtained from
the generalised Maxwell relaxation spectrum of Eq. \ref{Eq: General GM relaxation spectrum} (top) and
the generalised Kelvin-Voigt retardation spectrum of Eq. \ref{Eq: General GK retardation spectrum} (bottom),
with $\tau_{\min}=10^{-3}$, $\tau_{\max}=10^3$ (in arbitrary units), and $N=10^3$.
These exponents were computed numerically, for every $\alpha$, in the interior range $10^{-1}<w<10^1$.}
\label{fig3}
\end{center}
\end{figure}

As it may be seen from this figure, 
we recover the global behaviour previously described in the analysis of the continuous relaxation spectrum.
However, the referred power law exponents (including those of Eq. (\ref{Eq: Weak power law for complex modulus})) 
correspond only to an approximation
(they would be valid for $\tau_{\min}\to 0$ and $\tau_{\max}\to \infty$),
we show in Fig. \ref{fig3} the numerically converged (we have increased $N$ to verify it) computed values, 
established in the interior range $10^{-1}<w<10^1$.
It is important to mention that the exponent of $G'$ is always greater than the exponent of $G''$, even in the region $0<\alpha<1$,
where they assume almost identical values. This is a distinctive feature of the generalised Maxwell model.

It should be noted that we have performed the same calculations for different numbers of elements.
The possibility of changing $\beta$ allows us to have a good approximation of the power law behaviours even for
a reduced number of elements ($N>10$). Nevertheless, as the number of elements diminishes, some oscillations
of the elastic and viscous moduli start to appear.

\section{Generalised Kelvin-Voigt model}

We now turn to the generalised Kelvin-Voigt model (see Fig. \ref{fig1}, bottom) composed of a serial array of $N$ Kelvin-Voigt viscoelastic elements.
Each Kelvin-Voigt element is composed of a dashpot of drag coefficient $\eta_i$ in parallel with a spring of stiffness $G_i$.
It behaves elastically on long times scales and its dynamics comes from the viscous element.
Its creep compliance is $J_i(t)=J_i(1-e^{-t/\tau_i})$, where $J_i=1/G_i$ and $\tau_i=\eta_i/G_i$ is the element retardation time.
Because the Kelvin-Voigt elements are in series, the total creep compliance is just the sum:
\begin{equation}
J(t)=\sum_{i=1}^{N}J_i(1-e^{-t/\tau_i})
\label{Eq: GK creep compliance}
\end{equation}
This sum may be approximated by the integral:
\begin{equation}
J\approx \int_{\tau_1}^{\tau_N} J_\tau (1-e^{-t/\tau_i})\frac{di}{d\tau}d\tau
\end{equation}
If we want to establish the weak power law dependence $J_{GK}\sim t^\alpha$,
the retardation spectrum must then scale as
\begin{equation}
{\cal L}(\tau)\equiv J_\tau\frac{di}{d\tau}\tau\sim\tau^{\alpha}
\label{Eq: GK retardation spectrum}
\end{equation}
As before, the exponent $\alpha$ of the retardation spectrum may assume any value.

To understand the rheological behaviour of a material with a general power law spectrum of retardation times, 
we will determine the corresponding elastic and viscous moduli.
After performing a Laplace transform and using relations (\ref{Eq: Viscoelastic relations}), we first obtain
\begin{align}
\label{Eq: GK Elastic compliance}
J'(w)&\sim \int_{\tau_{\min}}^{\tau_{\max}} \tau^{\alpha}\frac{1}{1+(w\tau)^2}\frac{d\tau}{\tau} \\
J''(w)&\sim \int_{\tau_{\min}}^{\tau_{\max}} \tau^{\alpha}\frac{w\tau}{1+(w\tau)^2}\frac{d\tau}{\tau} 
\label{Eq: GK Viscous compliance}
\end{align}
The elastic and viscous moduli may then be determined from the relations:
\begin{align}
\label{Eq: G from J}
G'=\frac{J'}{J'^2+J''^2},\;\;\;
G''=\frac{J''}{J'^2+J''^2}
\end{align}

For $w \ll 1/\tau_{\max}$, the integrals of Eq. \ref{Eq: GK Elastic compliance} and 
\ref{Eq: GK Viscous compliance}
are much simplified, and we obtain the scalings $J'\sim w^0$ and $J''\sim w^1$, with $J'\gg J''$. 
The elastic and viscous moduli scale as $G'\approx 1/J'\sim w^0$ and $G''\approx J''/J'^2\sim w^1$.
For $w \gg 1/\tau_{\min}$, the same integrals are again simplified, 
yielding the scalings $J'\sim w^{-2}$ and $J''\sim w^{-1}$, with $J'\ll J''$. 
We obtain then the same scalings for the elastic and viscous moduli: $G'\approx J'/J''^2\sim w^0$ and $G''\approx 1/J''\sim w^1$.
The coefficients of $G'$ and $G''$ are dependent of $\alpha$ but their scalings are not. In fact, the exponents of $w$ 
coincide with the scalings of a simple Kelvin-Voigt model.

For $1/\tau_{\max}\ll w \ll 1/\tau_{\min}$, we may in some cases extend the limits of the integrals of 
Eq. (\ref{Eq: GK Elastic compliance}) and 
(\ref{Eq: GK Viscous compliance})
to $\tau_{\min}\to 0$ and $\tau_{\max}\to\infty$, which allows us to obtain the results:
\begin{align}
\label{eq elastic compliance 2}
J'(w)&\sim \frac{\pi}{2} w^{-\alpha}\text{csc}\frac{\pi \alpha}{2}\;\;\;(\text{if}\;\;0<\alpha<2) \\
J''(w)&\sim \frac{\pi}{2} w^{-\alpha}\text{sec}\frac{\pi \alpha}{2}\;\;\;(\text{if}\;\;-1<\alpha<1) 
\label{eq viscous compliance 2}
\end{align}
When $0<\alpha<1$, both integrals are well defined and we recover 
the weak power law behaviours of Eq. (\ref{Eq: Weak power law for complex modulus}).

If $\alpha<0$, the integral for the elastic compliance (Eq. (\ref{Eq: GK Elastic compliance})) diverges as $\tau_{\min}\to 0$.
The elastic compliance is then dominated by the smallest retardation time. 
In this case, we have $w\tau_n\ll 1$, and $J'\sim w^0$.
On the contrary, if $\alpha>2$, then this integral is dominated by the largest retardation time, $\tau_{\max}$. 
We have $w\tau_1\gg 1$, and the power law behaviour $J'\sim w^{-2}$.
By the same line of reasoning, we may determine from Eq. (\ref{Eq: GK Viscous compliance})
the power law behaviours $J''\sim w^1$ for $\alpha<-1$ and $J''\sim w^{-1}$ for $\alpha>1$.
Applying these results to each interval of $\alpha$, and using the approximations $J'\gg J''$ for $\alpha<0$ and $J'\ll J''$ for $\alpha>1$
(which we may infer from Eq. (\ref{eq elastic compliance 2}) and (\ref{eq viscous compliance 2})), we obtain the power law behaviours:
\begin{align}
&G'\sim w^0 & &G''\sim w^1 & &(\alpha<-1) \\
&G'\sim w^0 & &G''\sim w^{-\alpha} & &(-1<\alpha<0) \\
&G'\sim w^\alpha & &G''\sim w^{\alpha} & &(0<\alpha<1) \\
&G'\sim w^{1-\alpha} & &G''\sim w^1 & &(1<\alpha<2) \\
&G'\sim w^0 & &G''\sim w^1 & &(2<\alpha)
\end{align}
We note that for $\alpha<-1$ or $\alpha>2$, we have a single power law behaviour for $G'\sim w^0$ and $G''\sim w^1$, for all values of the angular frequency $w$.
Indeed, for these ranges of $\alpha$, the whole structure is entirely dominated by only one Kelvin-Voigt element, 
corresponding respectively to the minimum ($\alpha<-1$) or the maximum ($\alpha>2$) retardation times.

To plot representative elastic and viscous moduli {\it vs} $w$, for different values of $\alpha$, we used the discrete generalised Kelvin-Voigt model, with $N$ well defined retardation modes.
Again, the power law retardation spectrum of Eq. (\ref{Eq: GK retardation spectrum}) may be implemented by
two independent functions, $J_\tau=1/G_\tau$ and $di/d\tau$. So, one of them may be chosen arbitrarily.

If we choose the power law behaviour $J_i=Ji^{\beta}$
the condition given by Eq. \ref{Eq: GK retardation spectrum} implies the following discrete spectrum of retardation times:
\begin{equation}
\tau_i=\tau_1i^{(1+\beta)/\alpha}\;\;\;\;(J_i=Ji^\beta)
\label{Eq: General GK retardation spectrum}
\end{equation}
For a given set of $\tau_{\max}$, $\tau_{\min}$, $\alpha$, we are able to construct the weak power law behaviours of 
Eq. (\ref{Eq: Weak power law for complex modulus})
from a generalised Kelvin-Voigt model with $N$ modes by making $\tau_1=\tau_{\min}$ and 
$\beta=( \alpha \ln (\tau_{\max}/\tau_{\min})/\ln N)-1$.

The exact elastic and viscous compliances are given through the Laplace transform of Eq. (\ref{Eq: GK creep compliance}):
\begin{align}
\label{Eq: J' sums}
J'(w)&=\sum_{i=1}^{N}J_i\frac{1}{1+(w\tau_i)^2}\\
J''(w)&=\sum_{i=1}^{N}J_i\frac{w\tau_i}{1+(w\tau_i)^2}
\label{Eq: J'' sums}
\end{align}
In order to determine the elastic and viscous moduli we use Eq. (\ref{Eq: G from J}).

Figure \ref{fig2} (right) shows the elastic and viscous moduli, $G'_{GK}(w)$ and $G''_{GK}(w)$ for $\alpha=-0.5,0.25,0.5,0.75,1.5$,
$\tau_{\min}=10^{-3}$, $\tau_{\max}=10^3$ (in arbitrary units), and $N=10^3$.
To obtain them, we have used the sums of Eq. (\ref{Eq: J' sums}) and (\ref{Eq: J'' sums}), together with
the discrete retardation spectrum defined by Eq. (\ref{Eq: General GK retardation spectrum}).

As it may be seen, we recover the global behaviour previously described in the analysis of the continuous retardation spectrum.
As these power law exponents correspond only to an approximation, only valid as $\tau_{\min}\to 0$ and $\tau_{\max}\to\infty$,
we show in Fig. \ref{fig3} the numerically converged computed values, established in the interior range $10^{-1}<w<10^1$.
We note that the exponent of $G''$ is always greater than the exponent of $G'$. 
This is a distinctive feature of the generalised Kelvin-Voigt model.
This inversion in the values of the slopes is reflected also in the intersection points of $G'$ and $G''$ (see Fig. \ref{fig2}, right and left). If the intersection takes place at small values of $w$, for the generalised Maxwell model (when $\alpha<0.5$), it occurs
at large values of $w$ for the generalised Kelvin-Voigt model (for the same value of $\alpha$) and vice versa.

\section{Conclusion}

In this article, we have focused on the interesting weak power law behaviours $G'(w)\sim G''(w)\sim w^\alpha$ ($0<\alpha<1$) that occur
in a large variety of soft materials. 
They may appear in diluted, fluid systems, encompassing both the Brownian motion of elastic chains in a solvent,
modelling unentangled polymers, or the important SGM model, that describes fluid behaviour above the glass transition.
The weak power law behaviours may also occur in gelled, elastic systems, made out of polymeric networks, which 
retain two important characteristics of the cell's cytoskeleton, namely the idea of an
existing pre-stress on its fibres,
and the idea of a possible fractal structure, characterised by a power law distribution of elastic element lengths and retardation times \cite{patricio2015rheology}.
The cytoskeleton is however a very complex active structure, and exhibits other important rheological properties
that are not considered here, such as stress stiffening \cite{fernandez2006master} or a new power law exponent (of $3/4$) for higher frequencies, usually associated with the rheological behaviour of diluted semi-flexible filaments \cite{deng2006fast}.

Particularly interesting, we find that beyond the region corresponding to the weak power law behaviours ($\alpha<0$ or $\alpha>1$),
the generalised Maxwell and Kelvin-Voigt models give very different and distinctive rheological behaviours,
converging continuously to the 1-element Maxwell and Kelvin-Voigt models, respectively, for $\alpha<-1$ or $\alpha>2$.

\section*{Acknowledgements}

We thank C. R. Leal, J. Duarte, C. Janu\'ario, J. M. Tavares and P. I. C.Teixeira for stimulating discussions.
We thank P. I. C. Teixeira for linguistic assistance.

\bibliography{rheology}

\end{document}